\documentclass{INTERSPEECH2023}

\interspeechcameraready

\title{On the Adaptability and Robustness of Whisper}
\title{Systematic Evaluation of Tuned Whisper in Real Scenarios}
\title{ Evaluation of  Whisper on Downstream Speech Tasks}
\title{On the Transferability of Whisper-based Representations for ``In-the-Wild'' Cross-Task Downstream Speech Applications}

\name{Vamsikrishna Chemudupati$^1$, Marzieh Tahaei$^1$, Heitor Guimarães$^2$, Arthur Pimentel$^2$,\\ Anderson Avila$^2$, Mehdi Rezagholizadeh$^1$, Boxing Chen$^1$, Tiago Falk$^2$}
\address{
  $^1$Huawei Noah’s Ark Lab, Canada\\
  $^2$Institut National de la Recherche Scientifique, Canada}
\email{\{vamsikrishna.chemudupati,marzieh.tahaei,mehdi.rezagholizadeh,boxing.chen\}@huawei.com\\
\{heitor.guimaraes,arthur.pimentel,anderson.avila,tiago.falk\}@inrs.ca}

\begin{document}

\maketitle
 
\begin{abstract}
Large self-supervised pre-trained speech models have achieved remarkable success across various speech-processing tasks. The self-supervised training of these models leads to universal speech representations that can be used for different downstream tasks, ranging from automatic speech recognition (ASR) to speaker identification. Recently, Whisper, a transformer-based model was proposed and trained on large amount of weakly supervised data for ASR; it outperformed several state-of-the-art self-supervised models. Given the superiority of Whisper for ASR, in this paper we explore the transferability of the representation for four other speech tasks in SUPERB benchmark. Moreover, we explore the robustness of Whisper representation for ``in the wild'' tasks where speech is corrupted by environment noise and room reverberation. Experimental results show Whisper achieves promising results across tasks and environmental conditions, thus showing potential for cross-task real-world deployment.

\end{abstract}








\section{Introduction}

Self-supervised learning (SSL) has shown promising results in the field of speech processing, where a relatively large amount of unlabelled data is used as input to large models to learn universal representations that can be utilized on a variety of downstream tasks \cite{schneider2019wav2vec,baevski2020wav2vec,hsu2021hubert,huang2022spiral,chen2022wavlm}. In fact, recent works have shown that a weighted, task-specific sum of embeddings
from different layers of these models can be used as a powerful representation for fine-tuning on downstream tasks \cite{yang2021superb}. However, as shown by the results in \cite{chen2022wavlm}, these models perform well on downstream automatic speech recognition (ASR) tasks, but their performance on scenarios that involve multiple speakers is not satisfactory. These findings suggest that the representations optimized for ASR likely discard speaker-related characteristics to focus on robust linguistic content recognition. In turn, for e.g. in speaker identification, representations should focus more on speaker-related information and less on linguistic content. Moreover, unless specific strategies are put in place, existing universal representations face significant performance drops in the presence of environmental artifacts, such as background noise and room reverberation \cite{guimaraes2022improving}. 


One such strategy comprises training the models on a more diverse dataset with varying noise levels. The WavLM model \cite{chen2022wavlm}, for example, showed significant improvements in performance across various downstream tasks (e.g., speaker diarization, speaker identification, keyword spotting and speaker verification) by pre-training on speech data obtained from multiple domains and audiobooks. 
More recently, authors in \cite{radford2022whisper} proposed the so-called Whisper, a transformer-based model trained on massive amounts of data in a supervised manner. More specifically, training data consisted of 680,000 hours of weakly supervised data obtained from different environments, recording setups, speakers and languages. The model, therefore, is exposed to a mixed distribution of speech data and is forced to extract a representation that should generalize well. The model leverages multi-task training involving voice activity detection, speech recognition, language identification and translation and it was shown to outperform several state-of-the-art (SOTA) SSL models on an ASR task. The model also showed to be robust to ambient noise with signal-to-noise ratios ranging from 5 dB to 20 dB. In a zero-shot setup, the performance obtained was equivalent to that of a fine-tuned SSL speech models. 

Notwithstanding, the multi-task training paradigms inherent to Whisper are closely coupled to linguistic tasks, such as language and speech recognition. As such, it is not clear if the Whisper representation could be useful for other speaker or para-linguistic downstream tasks, such as speaker identification or emotion recognition. This paper aims to answer this research question. Moreover, we wish to explore if the noise robustness properties seen with ASR tasks \cite{radford2022whisper} remain with these other tasks. More specifically, we investigate the transferability of the Whisper robustness to four other downstream tasks in the SUPERB benchmark, namely: speaker identification, keyword spotting, intent classification, and emotion recognition. We do this under varying noise and reverberation conditions. Experimental results show that Whisper achieves promising results across tasks and environmental conditions.

\section{Related work}

In recent years, there has been a growing interest in self-supervised  training of speech models. In these models, the speech signal is often perturbed and  masked language modeling is utilized to learn contextualized speech representations.
The architecture of SSL models typically consists of a feature encoder using 1-dimensional convolution layers, followed by a transformer network to extract the contextualized representations from the high-level encoder features. The final representations are then taken as the predicted labels and used to compute the objective function along with the target labels. Models are usually differentiated based on the method of generating the target labels which are used to perform SSL on unlabelled data.

\begin{figure*}
   \centering
   \includegraphics[width=0.85\linewidth]{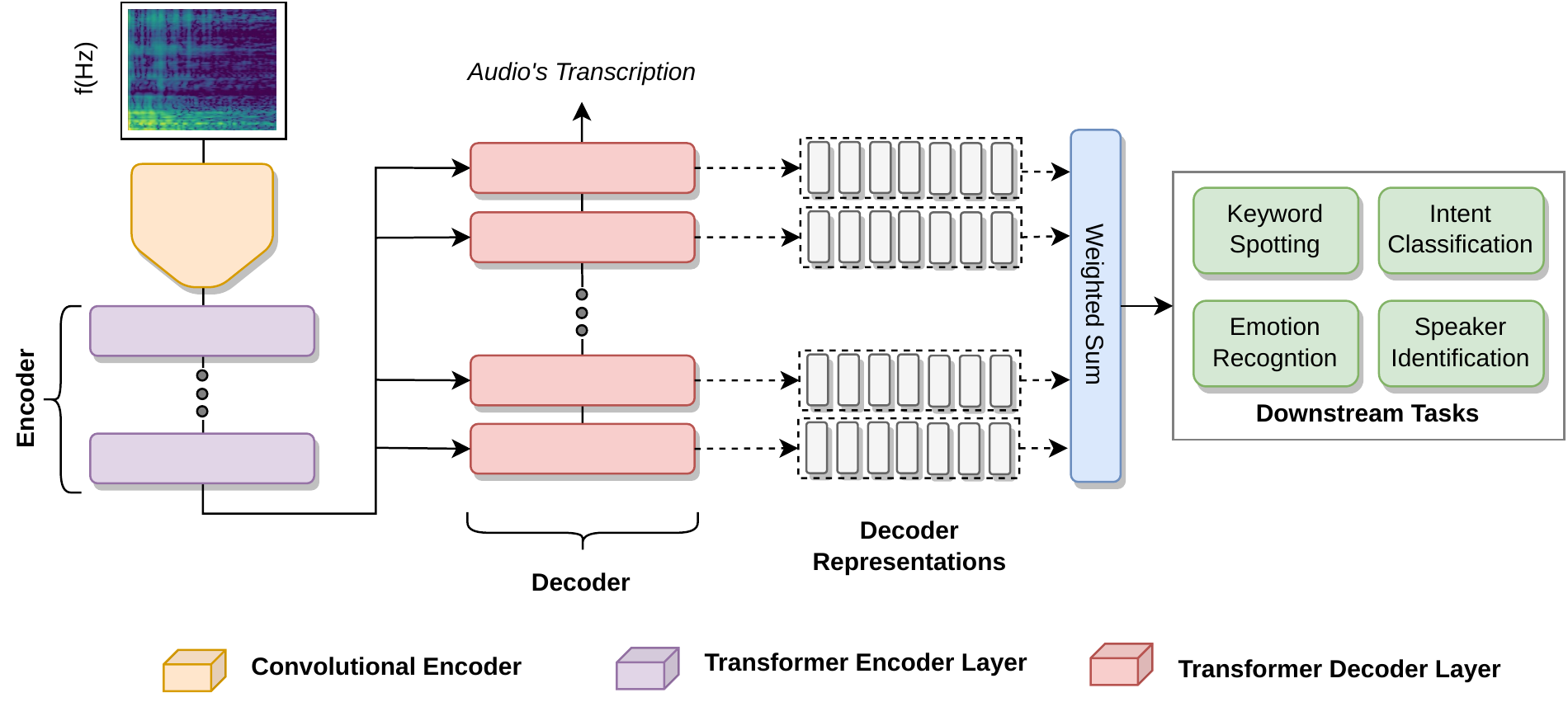}
   \caption{Diagram of the proposed cross-task experiments on top of Whisper.}
   \label{fig:noise_env}
\end{figure*}



Next, we provide a brief overview of some popular SSL models used within the speech community. The models will be used as baselines in our experiments:\\[4pt]
\textbf{Wav2vec 2.0} \cite{schneider2019wav2vec} proposes a method of generating target labels by converting the continuous speech representations to discrete form. The raw audio is encoded using the multi-layer convolutional neural network,and spans of the output representation are masked, and input to a transformer network in order to build contextualized representations. To perform self-supervised training, the output of the convolutional feature encoder must be discretized, which is achieved via product quantization \cite{jegou2010product}. Contrastive loss is used as the objective function to distinguish the true latent representation from distractors \cite{baevski2020wav2vec}.\\[4pt]

\noindent \textbf{HuBERT} \cite{hsu2021hubert} uses an unsupervised ensemble clustering approach to generate the targets to learn the representations. The architecture of HuBERT is similar to wav2vec 2.0. However, the authors propose an improved discretization step that aims at finding clusters that represent acoustic units (e.g., phonemes). Thus, the concept of hidden sound unit is introduced which is based on an offline multiple clustering step to improve the target quality. This step generates noisy labels for a BERT-like pre-training (i.e., based on predicted masked inputs). The predictive loss is only considered for masked regions. This forces the model to learn how to extract good high-level representations, based on unmasked inputs, to correctly predict the targets of masked ones. Given the input utterance, $X = [x_1, ..., x_T]$, the hidden units are denoted by $Z = [z_1, ..., z_T]$, where $h$ is the clustering function. A masked prediction model is defined as $p_f(\cdot , \tilde{X},t)$, which represents the distribution over target indices for each time step. The following cross-entropy loss is adopted to compute over masked predictions,

\begin{equation}
    L_m(f; X, M, Z) = \sum_{t\in M}p_f(z_t, \tilde{X},t)
\end{equation}

\noindent where M is the set of indices to be masked from $X$.\\[4pt]
\textbf{WavLM} \cite{chen2022wavlm} achieved SOTA performance by following a different methodology when compared to the above methods. The model emphasizes two novel directions for SSL training: (1) the use of a diverse dataset of 94,000 hours of audio having different types of noise and speakers and (2) for pre-training, the inputs are perturbed with simulated noisy/overlapped
speech followed by masking. As for the training objective, similar to HuBERT, the model tries to predict the label
of the original clean speech on the masked region.
This improves its potential for non-ASR tasks. It also uses a gated relative position bias on the transformer to help better extract sequential features in audio inputs. The training procedure is similar to HuBERT where a clustering approach is followed to generate the target labels. \\[4pt]
\textbf{Robust HuBERT} \cite{huang22b_interspeech}, in turn, is an innovation on top of the HuBERT recipe to create a robust model by leveraging domain adaptation techniques. While it is possible to improve data diversity through data augmentation, this may not be scalable since we can only observe some variation factors. To alleviate this problem, Robust HuBERT proposes to use domain adversarial training (DAT)~\cite{JMLR:v17:15-239} to learn noise-invariant features using the HuBERT model as a baseline. The domain classifier must distinguish if a latent representation comes from a clean or noisy sample while the model generates noise-invariant features.

To standardize cross-task experiments, the SUPERB (Speech processing Universal PERformance Benchmark) \cite{yang2021superb} was established. 
The benchmark helps evaluate whether new representations are simple or universal and can be used across numerous tasks. As fine-tuning pre-trained models can be computationally intensive, SUPERB scores are calculated using frozen pre-trained models and lightweight prediction heads. The benchmark includes ten tasks to evaluate the following aspects of speech: content, speaker, semantics, and para-linguistics. All datasets used are publicly available and have enough labeled data to benchmark pre-trained models. In this work, we experiment with four SUPERB tasks focusing on content, speaker, semantics and para-linguistics aspects. These tasks correspond to keyword spotting, intent classification, emotion recognition, and speaker identification.


\section{Whisper Model}
Whisper is a weakly supervised pre-trained speech recognition model which uses a attention-based encoder-decoder architecture, as shown in Figure~\ref{fig:noise_env}. 
For training, the audio is re-sampled at 16 kHz and a log-magnitude mel spectrogram representation is computed over 25 ms windows with a stride of 10 ms. The encoder uses two 1d-convolution layers and a GELU activation function to process the spectrogram frames and extract the high-level representations. Positional embeddings are added to the output representations and input to the transformer models. The encoder and decoder use the same hidden size and number of transformer blocks.

While the original Whisper model was trained on different sizes to study the scaling laws, here all experiments are performed using the base Whisper model with the following configuration: 74M parameters, 6 layers, 512 hidden size, and 8 attention heads. The Whisper model is trained using two methodologies: (1) Multi-task learning where the model has been trained on multiple tasks, such as English transcription, Any-to-English transcription, and non-English transcription, and  (2) trained only on English language data and performed a single task of English transcription. We use the English language Whisper model for experimentation as the SUPERB datasets used for fine-tuning support primarily English language.

\begin{table*}
    \centering
    \caption{Performance comparison for  Keyword Spotting (KS), Intent Classification (IC), Emotion Recognition (ER) and Speaker Identification (SID) under clean conditions.}
    \label{tab:results1}
    \begin{tabular}{cccccccc}
        \toprule
        Upstream & \#Params  & Pre-training Corpus & KS & IC & ER & SID & Overall\\
        \midrule
        Wav2vec 2.0 Base & 95M & LS 960 hr & 96.23 & 92.35 & 63.43 & 75.18 & 81.79 \\
        HuBERT Base & 95M & LS 960 hr &96.30 & 98.34  & 64.92 & 81.42 & 85.24 \\
        WavLM Base & 94.7M & LS 960 hr & 96.79 & 98.63 & 65.94& 84.51 & 85.36 \\
        WavLM Base+ & 97M & Mix 94k hr &  97.37 & 99.00 & 68.65& \textbf{89.42} & \textbf{88.61} \\
        \midrule
        Whisper Base &  74M & 680K hr (labeled)  & \textbf{97.63} & 95.17  & 64.96 & 7.57  
  & 65.92 \\
  Whisper Base (fine-tuned) &  74M & 680K hr (labeled)  & 95.58
 & \textbf{99.45}  & \textbf{68.87} & 84.85
  & 87.18\\

    \midrule[\heavyrulewidth]
        \bottomrule
    \end{tabular} 
    
\end{table*}
\section{Experimental Setup}
Here, we describe the downstream tasks, the testing setup, datasets used, and model training.

\subsection{Downstream tasks and robustness evaluation}

To gauge the generalizability and robustness of Whisper, we investigate a subset of four downstream tasks from the original SUPERB \cite{yang2021superb}, namely Keyword Spotting (KS), Intent Classification (IC), Emotion Recognition (ER) and Speaker Identification (SID). The four tasks are evaluated under the standard accuracy metric. We do not fine-tune the model on the ASR task as this is the primary function of Whisper and we wish to explore its use for applications other than its original intent.

Keyword spotting consists of a twelve-class classification problem where each utterance needs to be classified into one of 10 predetermined commands, silence, or unknown. Similarly, the intent classification task assigns labels to utterances using the following classes: action, object, and location. Next, emotion recognition is a multi-class classification problem where we assign one of the possible labels to each utterance: neutral, happy, sad, or angry. Finally, the speaker identification task, as the name suggests, aims to identify the speaker's identity for each utterance from a set of fixed speakers. The interested reader is referred to \cite{yang2021superb} for more details on the SUPERB tasks.

Here, we used the same approach as in \cite{guimaraes2023robustdistiller} to corrupt the test data with typical noise conditions seen in ``in-the-wild'' settings. Overall, models are tested across four scenarios: (1) clean, $(c)$, which is the same as the SUPERB tasks and uses the datasets as is; (2) noise-only $(n)$, where background noise is added to each utterance at an SNR ranging uniformly from -5 dB to 20 dB; (3) reverberation-only, $(r)$, where room acoustic effects are generated by convolving the speech signal with room impulse responses of varying sizes simulating small, medium, and large-sized rooms; and (4) noise-plus-reverberation, $(n+r)$, where both additive and convolutive noise are simultaneously applied to each utterance. More details about the noise datasets are given in Section 4.2.

\begin{table*}
    \centering
    \caption{The performance of Whisper and baseline models under noisy (\textbf{n}), reverberation (\textbf{r}), and noise-plus-reveberation (\textbf{n+r}) conditions. The relative percentage drop with respect to the clean condition are reported for easier comparison.}
    \label{tab:results2}
    \resizebox{\textwidth}{!}{%
    \begin{tabular}{ccccccccccc}
        \toprule
        & \multicolumn{3}{c}{KS} & \multicolumn{3}{c}{IC} & \multicolumn{3}{c}{ER} & \multirow{2}{*}{\textbf{Overall}} \\

        \cmidrule(lr){2-4}
        \cmidrule(lr){5-7}
        \cmidrule(lr){8-10}

        Upstream & (n)$~|~$drop & (r)$~|~$drop & (n + r)$~|~$drop & (n)$~|~$drop & (r)$~|~$drop & (n + r)$~|~$drop & (n)$~|~$drop & (r)$~|~$drop & (n + r)$~|~$drop &   \\ 

        \midrule
        wav2vec2 Base &86.50$~|~$10.08&64.43$~|~$33.02&56.25$~|~$41.52&69.26$~|~$25.39&68.78$~|~$25.90&50.88$~|~$45.19&51.46$~|~$18.33&32.70$~|~$48.10&27.02$~|~$57.11&56.36 \\
        HuBERT Base &84.55$~|~$12.20&61.70$~|~$35.92&51.61$~|~$46.40&78.70$~|~$19.97&75.77$~|~$22.95&59.08$~|~$39.92&53.45$~|~$17.45&40.72$~|~$37.11&29.12$~|~$55.02&59.41 \\
        Robust HuBERT &92.15$~|~$4.33&74.81$~|~$22.33&66.50$~|~$30.96&91.43$~|~$7.32&85.32$~|~$13.52&71.10$~|~$27.93 & 57.17$~|~$11.48&41.44$~|~$35.84&33.36$~|~$48.35&68.14 \\
        WavLM Base+&90.07$~|~$7.02&78.64$~|~$18.82&72.38$~|~$25.28&83.13$~|~$15.89&\textbf{89.35}$~|~$\textbf{9.60}&68.20$~|~$30.99&55.08$~|~$18.95&36.57$~|~$46.18&27.14$~|~$60.06&66.72 \\
        
        \midrule
        Whisper Base&\textbf{96.98}$~|~$\textbf{0.66}&\textbf{89.58}$~|~$\textbf{8.24}&\textbf{88.18}$~|~$\textbf{9.67}&\textbf{88.74}$~|~$\textbf{6.75}&85.05$~|~$10.63&\textbf{80.88}$~|~$\textbf{15.01}&\textbf{64.15}$~|~$\textbf{1.24}&\textbf{45.15}$~|~$\textbf{30.49}&\textbf{38.19}$~|~$\textbf{41.20}&\textbf{75.21} \\

    \midrule[\heavyrulewidth]
        \bottomrule
    \end{tabular} \vspace{-6mm}
     }
\end{table*}

\subsection{Datasets}
In this paper, the wav2vec 2.0 and HuBERT benchmark models are trained on the LibriSpeech (LS) corpus \cite{panayotov2015librispeech}, which contains 960 hours of audiobook recordings. While the WavLM benchmark model is trained on the Mix 94k hr dataset, which consists of 60k hours of Libri-Light \cite{kahn2020libri}, 10k hours of GigaSpeech \cite{Chen2021}, and 24k hours of VoxPopuli \cite{Wang2021}. The WavLM model also uses the DNS4 noise dataset \cite{dubey2022icassp} in its training stage. The Whisper model, on the other hand, is trained on 680,000 hours of audio. Of those, 117,000 hours cover 96 languages other than English. The dataset also includes 125,000 hours of Any-to-English speech translation data \cite{radford2022whisper}. 

The datasets used to evaluate the downstream tasks are the same as the ones used in SUPERB. These are detailed below:
\begin{itemize}
\item{\textbf{KS}: The dataset used for fine-tuning is the Speech-commands v0.01\cite{Warden2018}. Train, test and validation data splits are 51,093, 6,799 and 3,081 audio files, respectively.}
\item{\textbf{IC}: The dataset used for fine-tuning is the Fluent speech commands \cite{lugosch2019speech,lugosch2020using}. The train, test and validation data splits are 23,132, 3,118 and 3,793 audio files, respectively, labeled with their corresponding intents.}
\item{\textbf{ER}: The dataset used for emotion recognition is IEMOCAP \cite{Busso2008} and it is split into five folds equally where four folds are used for training and one fold is used for testing.}
\item{\textbf{SID}:The dataset used is Voxceleb1 \cite{Nagrani_2017} which has 1,251 different speakers. The train and test splits for audio utterances are 145, 265 and 8,251 respectively.}
\end{itemize}

To evaluate the performance of our models under unseen conditions, we conduct testing using multiple noise datasets. We use the DCASE2020  \cite{Mesaros2018_DCASE} dataset, which contains 64 hours of audio recordings in 10 different acoustic scenes recorded in 12 different cities. For reverberation, we use the OpenSLR26  \cite{ko2017study} dataset, which contains 60,000 simulated room impulse responses. 
\subsection{Training setup}

For fine-tuning, input audio chunks of 30s are converted to log Mel spectrogram and given as input to the Whisper encoder. The whisper model is then used as the upstream model that extracts the high-level audio representations. The resulting representations are then fed to the downstream task module.  As for downstream modules we use the lightweight prediction heads provided by SUPERB benchmark.

Given that Whisper was trained using supervised training, freezing the upstream model as suggested by SUPERB may not lead to satisfactory results for all tasks. Therefore, we use two fine-tuning procedures: 1) the upstream whisper model is frozen and the weighted sum of the decoder hidden states are passed to the downstream layers for obtaining predictions. This method is used to examine the usefulness of universal representations and it is also followed for robustness experiments. 2) Here, the complete model is fine-tuned and the hidden states of the last decoder layer are passed to the downstream layers. The complete setup is illustrated in Figure~\ref{fig:noise_env}.

The hyper-parameter settings used are the default ones given in SUPERB and explained for each task below:
\begin{itemize}
\item{\textbf{KS}, \textbf{IC}, \textbf{SID}: We train with a batch size of 32 for 200,000 training steps at a learning rate of 1e-4 using the Adam optimizer.}
\item{\textbf{ER}: We perform cross-validation training, where 4 folds are used as train data and 1 fold as test data. The model is trained for 30,000 steps with a batch size of 32 at a learning rate of 1e-4 using the Adam optimizer.}
\end{itemize}
All the training experiments are performed with a single V100 GPU. Early stopping is used during the fine-tuning procedure to avoid over-fitting. 
For our robustness analysis fine-tuned checkpoint of SOTA models were needed. Therefor,
the baseline values in Table 2 are obtained by fine-tuning all SOTA models individually using the above-listed hyper-parameters, due to the unavailability of task-specific fine-tuned checkpoints.

\section{Experimental Results and Discussion}
In this section, we present the obtained findings and discuss them in light of existing literature.

\subsection{Cross-task transferability results}
Table 1 reports the obtained results from Whisper as it is fine-tuned for four different SUPERB tasks, alongside different benchmarks obtained from \cite{chen2022wavlm}. It is evident that Whisper achieved SOTA accuracy across three of the four tasks with 20M fewer parameters relative to the benchmarks. In turn, it is observed that Whisper's performance varies across tasks based on whether its weights are frozen or not during training. 

For example, for ASR-related tasks, such as KS, the Whisper base model achieves an accuracy of 97.63\% when the weights are frozen, thus outperforming the completely fine-tuned model by 2.5\%. For non-ASR tasks, in turn, such as IC, SID, and ER, the complete fine-tuned model performs better than the frozen model. The frozen model in SID has the worst performance among all non-ASR tasks, which shows the need for complete fine-tuning of Whisper in scenarios where the task adaptation requirement is high. Overall, the Whisper decoder model strategy trained on a diverse dataset has resulted in representations that seem to provide SOTA accuracy across tasks. The encoder-decoder architecture of Whisper, however, requires high computation and training time during the fine-tuning procedure, especially compared to other encoder-based only SSL models. As such, there is a trade-off between performance and computational requirements. The overall results suggest that Whisper's representations are not universal and cannot be directly used for non-ASR tasks. However complete fine-tuning for the same train steps as the SSL models helps in achieving results better than SOTA models for the non-ASR tasks, which demonstrates that weakly supervised training with a large-scale dataset helps in achieving generalized, re-usable and robust representations.

\subsection{Environment robustness evaluation}
Table 2 reports the results achieved with the Whisper model and the benchmarks for the noise scenarios. All the models used in this section are fine-tuned by freezing the upstream model's weights. We do not include SID in robustness experiments due to the poor performance of Whisper in the clean scenario which prevents us from evaluating it further. The performance drops relative to the clean conditions are also reported for comparison purposes. As can be seen, Whisper achieves the lowest performance drop across the majority of the tested scenarios, except for intent classification under the reverberation condition. In fact, whenever reverberation was present, the performance drop was more substantial, suggesting the sensitivity of the model to room acoustic effects. Overall, the drops were most substantial in the noise-plus-reverberation conditions, suggesting there may still be some room for improvements.

\section{Conclusion}
In this work, we investigate the transferability of the  Whisper model to non-ASR based tasks, as well as its robustness to ``in the wild'' conditions, including background additive noise, room reverberation, and their combined effects. We show that Whisper can achieve promising results when weights are kept frozen during training, achieving results in line with several SOTA benchmarks on three out of four tasks. Further performance improvements are seen when weights are fine-tuned, showing that Whisper model performs better than SOTA benchmarks on majority of the evaluated tasks on clean data. Moreover, our robustness evaluations show that Whisper model outperforms SOTA SSL models across a majority of tested noisy conditions on in-the-wild scenarios. Overall, the obtained results show the benefits of using large-scale datasets during pre-training, even when the pre-training objective is different from the downstream tasks.





\bibliographystyle{IEEEtran}
\bibliography{mybib}

\end{document}